\documentstyle[aps,prl,epsf]{revtex}
\begin{document}
\draft
\tighten
\twocolumn[\hsize\textwidth\columnwidth\hsize\csname
@twocolumnfalse\endcsname
\title{Intersubband magnetophonon resonances in quantum cascade
structures}
\author{D. Smirnov$^{a\dagger}$, O. Drachenko$^a$, J. Leotin$^{a\ddag}, \
$H. Page$^b$, C. Becker$^b$, C. Sirtori$^b$, V. Apalkov$^c$, 
T. Chakraborty$^{c\ast}$}
\address{$^a$Laboratoire National de Champs Magn\'{e}tiques
Puls\'{e}s et Laboratoire de Physique de la Mati\`{e}re 
Condens\'{e}e, \hfil\break
143 Avenue de Rangueil, 31432 Toulouse, France}
\address{$^b$ Laboratoire Central de Recherches Thal\`{e}s, 
91404 Orsay, France}
\address{$^c$ Max-Plank-Institut f\"ur Physik komplexer Systeme,
01187 Dresden, Germany}
\date{\today}
\maketitle
\begin{abstract}
We report on our magnetotransport measurements of GaAs/GaAlAs 
quantum cascade structures in a magnetic field of up to 62 T. We 
observe novel quantum oscillations in tunneling current that are 
periodic in reciprocal magnetic field. We explain these oscillations 
as intersubband magnetophonon resonance due to electron
relaxation by emission of either single optical or acoustic 
phonons. Our work also provides a non-optical {\it in situ} 
measurement of intersubband separations in quantum cascade structures.
\end{abstract}
\pacs{73.21.Fg,73.43.Qt,85.35.Be,42.55.Px}
\vskip2pc]
\narrowtext

Ever since the pioneering work on the magnetophonon effect by Gurevich 
and Firsov \cite{MPR}, high magnetic fields have been regarded as an 
important tool for investigation of the electron-optical-phonon 
interaction in semiconductor systems, particularly in confined 
structures. For the in-plane transport, quantization of the carrier 
motion in the plane into discrete Landau levels (LLs) of energies 
$(N+1/2)\hbar \omega_c$, $\omega_c=eB/m^{\ast}$ is the cyclotron 
frequency, gives rise to quantum oscillations at elevated 
temperatures due to resonant phonon absorption. These 
magnetophonon oscillations are periodic in $1/B$, and their 
strength is related to the electron-optical-phonon coupling, 
while the period gives either the effective mass or the energy 
of the participating optical phonons \cite{tsui}. On the other
hand, for perpendicular transport the magnetotunneling 
measurements in double barrier systems have allowed direct 
probing of the optical-phonon-assisted transitions from a 
quasi-two-dimensional (2D) emitter into empty LLs of the central 
well, as well as determining the effective mass carrier dynamics in 
these structures \cite{greg}. However, little is known from these double
barrier studies about intersubband relaxation via optical-phonon 
emission in quantum wells (QWs). A particularly interesting and 
unexplored situation occurs when the cyclotron energy exceeds the 
optical phonon energy and/or the subband energy separation. The 
first situation is achieved in a GaAs 2D electron gas at magnetic 
fields above 22 T \cite{tsui}, where the in-plane electron wave function is 
localized on a scale of the magnetic length, $l_c=\sqrt{\hbar/eB}$ 
(3.2 nm at 62 T), and the electron behavior is essentially zero 
dimensional.

Intersubband relaxation via optical phonon emission in quantum
wells plays a key role in intersubband radiation sources like
the quantum cascade lasers (QCLs) \cite{Faist}. These
systems consist of double-barrier-like structures with 
three subbands belonging to a central QW structure. When a high bias 
is applied to the QCL system, the upper subband is populated by 
tunneling injection.
The electron relaxation in the central wells is then essentially 
governed by the optical phonon emission rate from both upper
subbands. As we demonstrate below, the QCL is appropriate to study 
intersubband relaxation via optical phonon emission. Recently, 
intersubband relaxation
effects were observed in a GaAs/GaAlAs QC structure from both
magnetoresistance and and luminescence up to 8 tesla \cite{Gornik}. 
However, the QCL structure for these studies had intersubband 
separation much below the LO-phonon energy and therefore, was 
unable to show resonant relaxation due to optical phonons.

In this paper, we report on our tunneling magnetotransport 
measurements in a GaAs/GaAlAs QCL structure where a magnetic field 
of up to 62 T is applied parallel to the current. Our measurements 
indicate evidence of intersubband magnetophonon oscillations in 
the tunneling current. These novel oscillations are shown to 
originate from resonant intersubband relaxation of electrons
from the ground state of the upper subband of the central wells
into excited Landau level of the lower subbands. A 
remarkable feature of our results is the lack of dependence of 
magnetic field positions of these oscillations on the applied 
bias, as with the in-plane magnetophonon oscillations. These 
results are supported by theoretical calculations of 
relaxation rates through acoustic and optical phonons as a 
function of the magnetic field. An additional benefit of this study 
is a non-optical $in$ $situ$ measurement of intersubband separations 
in QCL structures.

\begin{figure}
\centerline{
\epsfxsize=2.6in
\epsfbox{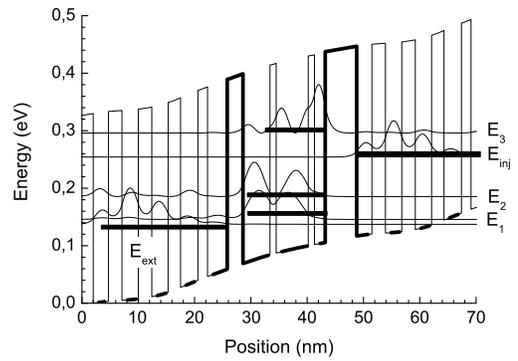}}
\vspace*{0.1in}
\protect\caption[QCL sample]
{\sloppy{ Electronic structure of a GaAs/GaAlAs QCL 
sequence with the double-barrier-like profile shown
by thick lines. The structure is biased at 117 mV per 
period. Ground state subband energy levels and their 
wave functions (squared) for the emitters on the right, 
central wells, and collectors on the left, are displayed. 
The layer thicknesses (in nm) along a period starting from 
the injector on the right toward the central well structure 
are: 3.6/{\bf{1.7}}/3.2/\underline{{\bf{2.2}}/3.0/{\bf{2.4}}/3.0}%
/{\bf{2.4}}/2.8/\frame{\bf{5.6}}/1.9/{\bf{1.1}}/%
5.8/{\bf{1.1}}\hfil\break/4.9/\frame{\bf{2.8}}. 
GaAlAs barriers are shown in bold and the rectangles indicate
barriers on both sides of the central well structure
with the three subbands $E_1, E_2$, and $E_3$. The underline
indicates a silicon doped region at a concentration of
6$\times10^{11}$ cm$^{-2}$. 
}}
\label{figone}
\end{figure}

The n-doped GaAs/GaAlAs QCL structure used for our measurements 
contains forty periods \cite{kruck}. Figure~\protect\ref{figone} 
shows the results of a self-consistent calculation of the electronic 
structure along a sequence of the biased QCL sample. The sequence
can be depicted like a double barrier tunneling structure 
including an injector/extractor with five graded gap wells, a 
wide injection barrier, a three well central zone, and a thin 
extraction barrier followed by the identical injector/extractor. 
The energy levels of the central wells, $E_i$ are indicated by thick 
horizontal lines in Fig.~\protect\ref{figone}. The emitter on 
the right hand side of the wide injection barrier behaves 
as a wide well developing a set of subbands labeled $E_n^{\rm inj},$ 
while the collector is a similar well on the left side of the 
extraction barrier. This picture identifies tunneling from 2D 
injector/extractor states into/out of 2D active zone states. 
On the other hand, when a magnetic field is applied perpendicular 
to the layers, tunneling takes place between 0D degenerate 
Landau states $E_{i,N}$, localized in a magnetic length scale $l_c$. 

\begin{figure}
\centerline{
\epsfxsize=3.0in
\epsfbox{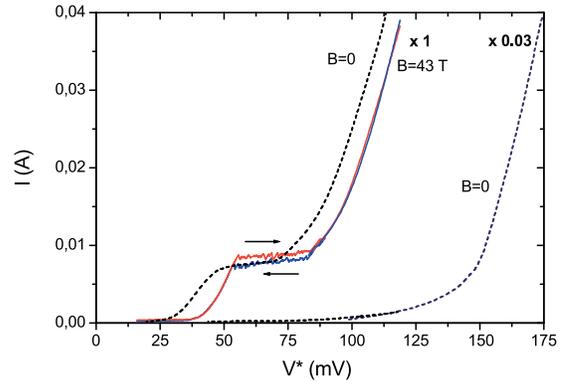}}
\vspace*{0.1in}
\protect\caption[figuretwo]
{\sloppy{$I(V^{\ast})$ curves measured at 4.2 K during the up and 
down voltage ramps at 0 T (dashed line) and 43 T (full line). The 
two curves correspond to two different scale values on the vertical
axis: one for the central curve, and 0.03 for the right curve. 
}}
\label{figtwo}
\end{figure}

Magnetotransport measurements were performed at 4.2 K on a sample 
processed into mesa-etched ridge waveguide bars, 20 $\mu$m wide and 
1.5 mm long. We used a pulsed magnet up to 62 tesla with total duration 
of 100 ms. The coil is based on a copper-stainless steel wire
developed at LNCMP, Toulouse \cite{coil}. We have performed mainly 
magneto-tunneling signal measurements under a fixed DC current or 
voltage bias. In the following, $V^{\ast}$ denotes voltage values 
across the QCL divided by the number of periods. Fast measurements 
of $I(V^{\ast})$ curves under quasi-static magnetic fields are also 
reported. The curves look globally the same when the magnetic field 
is changed from zero to 62 T as shown in Fig.~\protect\ref{figtwo}. 
The current rises in two steps separated by a plateau with a 
hysteresis behavior. Since the plateaus consist of multiple negative 
differential segments, this indicates a voltage range where the
tunneling transmission is decreasing after a resonant peak.
We therefore identify the edge of the plateau as resonant 
tunneling from $E_1^{\rm inj}$ into $E_{2,0}$ and the second 
steep rise as beginning of tunneling into $E_{3,0}$ 
level. The $I(V^{\ast})$ curve profile accounts clearly for the 
2D state tunneling.

\begin{figure}
\centerline{
\epsfxsize=3.0in
\epsfbox{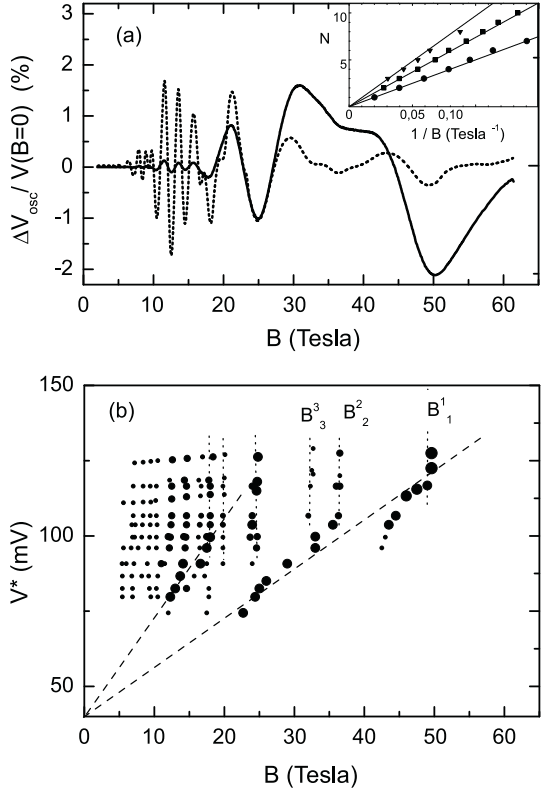}}
\vspace*{0.1in}
\protect\caption[figurethree]
{\sloppy{(a) Oscillatory component of the voltage across the 
QCL biased at 60 mA and measured in a field of up to 62 T.
The second derivative signal is drawn as the dotted line. The 
inset shows resonance numbers versus the inverse magnetic 
field, indicated by three series. (b) Oscillation extrema at a 
given bias. The dot size is related to the oscillation amplitude 
strength. One type of oscillations is independent of the applied 
bias, while the other changes roughly linearly with bias. The 
slope of the dashed line is equal to the cyclotron energy.
}}
\label{figthree}
  \end{figure}

In what follows, we investigate the quantum oscillations on 
longitudinal magnetotransport in the bias range well above the
hysteresis and current instabilities region. The quantum 
oscillation extrema are labeled as voltage minima under 
constant current bias or conversely, as current maxima under 
constant voltage bias. We limit the bias to the range below 
110 mA, where measurements can be made under DC bias without 
sample thermal drift. Figure~\protect\ref{figthree} (a) shows 
a typical recording of the oscillatory component of the
magnetoresistance of the structure up to 62 T. The second derivative 
curve is drawn as a dotted line. The inset shows resonance numbers 
versus the inverse magnetic field given by three curves. 

Figure~\protect\ref{figthree} (b) displays the magnetic field 
positions of oscillation extrema at measured voltage bias $V^*$
applied to the QCL structure. We observe two types of oscillations.
The first one occurs mainly in the lower range of bias $75-100$ mV
($12-35$ mA). In this case, oscillation field positions change
roughly linearly with bias as indicated by dashed lines.
The curves are drawn with cyclotron energy slopes according to the
relation, $eV^*=eV_0^*+N\hbar\omega_c$ for $m^*=0.07m_0$.
The cyclotron energy slope of the oscillation extrema pattern
indicates a tunneling transition from the injector ground state
Landau level toward a sequence of excited Landau levels in the 
central wells. This happens when the tunneling is assisted
by the emission of a LO phonon as observed in a magneto-tunneling 
study of a GaAs/AlGaAs double barrier structure \cite{greg}. 

Let is now consider the second type of oscillations observed
in the upper bias range above 100 mV. The oscillation field positions
do not change with bias as shown in Fig.~\protect\ref{figthree} (b).
Figure~\protect\ref{figthree} (a) shows in the oscillatory component 
of the voltage across the QCL structure that one series of peaks with the
largest amplitude occurs at a fundamental field near 50 tesla. The other 
two series are also visible on the second derivative curve. All
these series are identified in the inset of Fig.~\protect\ref{figthree} (a)
which plots integer $N$ versus the reciprocal magnetic field.
We label the harmonics in the three series as $B_N^i$ with $i=1,2$ and
3 indicating the lower, middle and upper curves respectively. The 
fundamental fields are obtained from the slopes at the 
values $B_1^1=50$ T, $B_1^2=75.5$ T, and $B_1^3=95$ T. We can then
derive the cyclotron energies $\varepsilon^{(i)}=\hbar eB_1^i/m^*$
by using the effective mass values measured in 2D GaAs electron gas
in the same field range as the fundamental fields \cite{miura}.
We obtain, $\varepsilon^{(1)}=72$ meV, $\varepsilon^{(2)}=109$ meV,
and $\varepsilon^{(3)}=149$ meV. Since these energy values are given by
the series that are insensitive to the QCL bias, they must be 
related to the electronic structure of the central wells. In fact, 
we find $\varepsilon^{(2)}$
and $\varepsilon^{(3)}$ are nearly equal to the photon energies
that we measured by the electroluminescence on this structure,
namely, 108.7 meV (laser line) and 147 meV (weak luminescence
line), respectively. We may then identify the energies $\varepsilon^{(2)}$
as $E_{3,0}-E_{2,0}$ and $\varepsilon^{(3)}$ as $E_{3,0}-E_{1,0}$.
The resonance condition for the two series (middle and upper
lines in the inset) are then expressed as: 
$E_{3,0}-E_{2,0}=N\hbar\omega_c$, and $E_{3,0}-E_{1,0}=N\hbar\omega_c$.
This corresponds to the elastic
transition from the $E_{3,0}$ into an excited Landau level of
the other subbands, $E_2$ and $E_1$, respectively.

On the other hand, if one adds the LO-phonon energy (36 meV) to 
$\varepsilon^{(1)}$, one gets 108 meV, which is close to $E_{3,0}-E_{2,0}
=108.7$ meV. Therefore, the resonance condition for the series
corresponding to the lower curve could be $E_{3,0}-E_{2,0}-\hbar
\omega_{LO}=N\hbar\omega_c$. This corresponds to
an inelastic transition with LO-phonon emission into an excited
Landau level of the $E_2$ subband. A similar inelastic transition
is expected into the $E_1$ subband when
$E_{3,0}-E_{1,0}-\hbar\omega_{LO}=N\hbar\omega_c$. In total, one
expects four series arising from elastic and inelastic intersubband
scattering. However, since in the QCL structure the separation
of the lower subbands equals the LO-phonon energy, the elastic
response transition from $E_{3,0}$ into $E_{2,N}$ subbands overlaps
with the inelastic resonance transition into $E_{1,N}$ subbands.
As a result, only three oscillation series should be obtained as
found in the experiment.

The observed tunneling current oscillations can indeed be accounted 
for by resonant intersubband relaxation of electrons with both 
optical and acoustic phonon emission. The optical phonon relaxation 
is possible only if the separation between
the levels $\Delta_i=E_3-E_i-N\hbar\omega_c=\hbar
\omega_{LO}$. For the acoustic phonons \cite{maksym} 
the resonance in electron relaxation rate occurs when
$\Delta_i=\hbar s/l_c$, where $s$ is the speed of sound.
$\Delta_i$ is about 1 meV at 60 T, which is much less than the 
intersubband separation.

The rate of electron transition from $E_{3,0}$ into $E_{i,N}$
subbands due to emission of LO  or acoustic phonons is 
\begin{eqnarray*}
\tau_{\mu,i}^{-1} &=&\frac{2\pi}{\hbar}\sum_j\int\frac{d\vec{Q}}{(2
\pi)^3}\delta(\Delta-\hbar\omega_{\mu}(Q))(1+n_{\mu j}(
\vec{Q}))\\
&\times&\left|M_{\mu,j}(\vec{Q})Z_i(q_z)\right|^{2}R_{0N}(q),
\end{eqnarray*}
where index $\mu$ stands for optical (LO) or acoustic (A) phonons, 
$n_{\mu j}(\vec{Q})$ is the phonon distribution function, 
$\vec{Q}=({\vec q},q_z)$ is the three dimensional vector, $j$ labels 
the phonon mode, $\omega_{LO}(Q)=\omega_{LO}$ and $\omega_A(Q)=sQ$. 
$M_{\mu j}(\vec{Q})$ are the matrix elements of electron-phonon interaction 
\cite{levinson}. For the acoustic phonons we took 
into account both deformation potential and piezoelectric couplings. 
The form factors $Z(q_z)$ and $R_{0N}(q)$ are given by the expressions 
\begin{eqnarray*}
Z_i(q_z) &=&\int dze^{iq_zz}\chi_3(z)\chi_i(z), \\
R_{0N}(q) &=&\frac1{N!}\frac{(ql_c)^{2N}}{2^{N}}e^{-(ql_c)^2/2}.
\end{eqnarray*}
To take into account the disorder effect, we introduce the broadening 
of Landau levels in a Lorentz form with the width of 2 meV and 
average the relaxation rate over Landau level distribution.

\begin{figure}
\centerline{
\epsfxsize=3.0in
\epsfbox{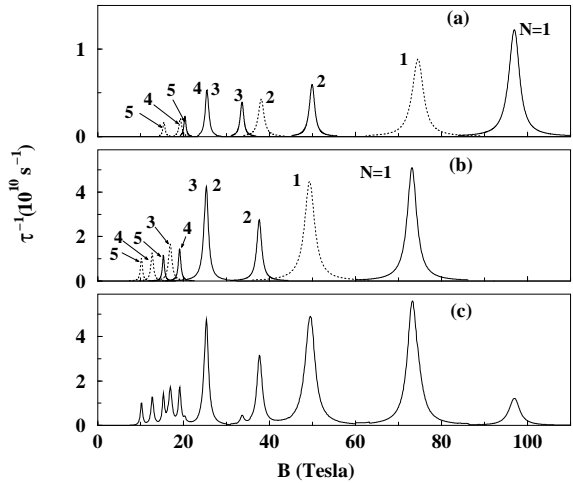}}
\vspace*{0.1in}
\protect\caption[figuresix]
{\sloppy{Acoustic-phonon (a) and LO phonon (b) emission rate 
as a function of the applied magnetic field. Electron 
transitions into $E_1$ and $E_2$ subbands are shown by 
solid and dotted lines, respectively. Lines marked by two numbers 
(4 and 3 in (a) and 3 and 2 in (b)) are the sum of two transitions: 
into $E_{1,4}$ and $E_{2,3}$ in (a) and $E_{1,3}$ and $E_{2,2}$ 
in (b). The total electron relaxation rate due to emission of 
acoustic or optical phonons as a function of magnetic field
is shown in (c).
}}
\label{Phonon}
  \end{figure}

In Fig.~\protect\ref{Phonon} (a), the total rate of acoustic 
phonon emission ($\tau_A^{-1}=\tau_{A,1}^{-1}+\tau_{A,2}^{-1}$) 
as a function of the magnetic field is shown for $N=1-5$ in units 
of $10^{10}$ s$^{-1}$. In the final state the electron is in 
$E_{1,N}$ or $E_{2,N}$ Landau ladder. There are sharp resonances 
when $E_3-E_2$ or $E_3-E_1$ is close to $N\hbar\omega_c$. 
Transitions to $E_{2,3}$ and $E_{1,4}$ are almost at the same 
magnetic field which is due to the particular structure of the 
QCL system: subband energies follow the relation, $E_3-E_2\approx
3(E_2-E_1)$, which result in double resonances when the 
energies of the levels $E_{2,3N}$ and $E_{1,4N}$ are close, 
where $N=1,2,..$.

In Fig.~\protect\ref{Phonon} (b), the total rate of LO phonon 
emission is shown as a function of the magnetic field for 
$N=1-5$. Due to the particular structure of the QCL system
($E_2-E_1\approx\hbar\omega_{LO}$), resonances due to optical 
phonons are almost at the same place as the resonaces due to 
acoustic phonons [Fig.~\protect\ref{Phonon} (a)]. Finally, in 
Fig.~\protect\ref{Phonon} (c) the total electron relaxation 
rate ($\tau^{-1}=\tau_A^{-1}+\tau_{LO}^{-1}$) due to emission 
of acoustic and optical phonons is shown. At $B\approx 25$T 
there are four resonances: two due to the acoustical 
phonons and the other two due to LO phonons. The model explains 
the observed resonance field positions fairly well. For example, 
the oscillation series for transitions to the $E_2$ subband 
with optical phonon emission agree within $\sim 1\%$ with the 
data for the $B_N^1$ series, and within $\sim3\%$ with the data 
for the $B_N^2$ series (transitions to the $E_1$ subband).

In summary, novel magnetophonon oscillation series were observed
in the tunneling current across a GaAs/GaAlAs QCL structure. In 
this system, which behaves like a double-barrier structure, the
oscillations originate from resonant intersubband relaxation 
from the upper subband inside the double-barrier central wells
by both optical and acoustic phonons. Magnetic field positions of these
oscillations are remarkably independent of the applied bias.
The observed results are supported by theoretical calculation of
relaxation rates through acoustic and optical phonons as a
function of the magnetic field.

We acklnowledge helpful discussions with Bruce McCombe.
We also acknowledge the cooperation of J. Galibert for the
pulsed field measurements and technical support from
R. Barbaste and C. Duprat.

\newpage


\end{document}